\documentclass{cpp2e}
\usepackage{epsfig, floatflt, rotating, latexsym, amssymb}
\newcommand{\be}{\begin{equation}}
\newcommand{\ee}{\end{equation}}
 \def\la{\;\raise0.3ex\hbox{$<$\kern-0.75em\raise-1.1ex\hbox{$\sim$}}\;}
 \def\ga{\;\raise0.3ex\hbox{$>$\kern-0.75em\raise-1.1ex\hbox{$\sim$}}\;}

\shorttitle{Strongly Coupled Plasma in Neutron Stars} 

\shortauthor{D.\ Yakovlev, O.\ Gnedin, A.\ Potekhin} 

\contvolum{41} 

\contyear{2001} 

\contnumber{no.\,2--3} 

\contpages{in press} 

\title{Strongly Coupled Coulomb and Nuclear Plasma in Inner Crusts
of Neutron Stars }

\author{D.G. Yakovlev$^{\rm a}$, O.Y. Gnedin$^{\rm b}$, 
A.Y. Potekhin$^{\rm a}$} 

\address{$^{\rm a}$ Ioffe Physical Technical Institute,
     194021, St.-Petersburg, Russia \\
	 $^{\rm b}$ Institute of Astronomy,
     Madingley Road, Cambridge CB3 0HA, England \\	
	 e-mail:yak@astro.ioffe.rssi.ru}	
\begin{document}
\setcounter{page}{1}
\makeheadings
\maketitle

\begin{abstract}
Matter of subnuclear density in
the inner crusts of neutron stars 
consists of neutron-rich atomic nuclei
immersed in strongly degenerate relativistic
gas of electrons and strongly nonideal liquid of
neutrons. Thermodynamic and kinetic
properties of this matter are greatly affected
by Coulomb and nuclear interactions and can be
studied, in principle, from observations of thermal
radiation of young (age $\la$ 100 yr) neutron stars.
\end{abstract}

\section{Introduction}
\label{introduc}

Neutron stars (NSs) are the most fascinating stars in the Universe.
Their masses are $M \sim 1.4 \, M_\odot$,
and their radii $R \sim 10$ km. Accordingly,
their mean mass density is about (2--3) $\rho_0$,
where $\rho_0 = 2.8 \times 10^{14}$ g cm$^{-3}$ is the mass
density of matter in atomic nuclei.
Thus, NSs
contain matter of supranuclear density compressed
by huge gravitational forces.
Accordingly,
they are treated as unique
astrophysical laboratories of such matter.
We show that young NSs are also laboratories
of strongly coupled Coulomb--nuclear plasmas of subnuclear density.

A NS consists of
a very thin atmosphere,
outer crust, inner crust, outer core and inner core 
(e.g., refs.\ \cite{st83, glendenning96}).
The outer crust is a few hundred meter thick extending
to the neutron drip density $\rho = \rho_{\rm d} 
\approx 4 \times 10^{11}$ g cm$^{-3}$. 
It mainly consists of strongly degenerate,
almost ideal electrons and fully ionized atoms (atomic nuclei).
The nuclei form a strongly nonideal Coulomb plasma,
liquid or solid, depending on $\rho$ and temperature $T$.
The properties of this matter are discussed in a companion
paper \cite{pcy00}.
The inner crust
(e.g., refs.\ \cite{nv73,oyamatsu93,pr95}), our main subject,
extends from $\rho=\rho_{\rm d}$ to
$\rho = \rho_{cc} \approx \rho_0/2$, and is about 1 km thick. 
Its matter
consists of electrons, neutron-rich nuclei,
and
Fermi liquid of neutrons dripped from the nuclei.
The neutrons are likely superfluid
due to the attractive part of nucleon-nucleon interaction. The critical
temperature of superfluidity 
is model dependent; its typical values range
from $10^9$ to $10^{10}$ K.
The nuclei arrange in a Coulomb
crystal; the nucleus shape and the parameters of neutron liquid
are governed by strong interaction.
The properties of this unique mixture of Coulomb
and strong interactions in matter of subnuclear density
are not well known from the theory.
They are especially uncertain at the crust 
bottom, $\rho \ga 10^{14}$
g cm$^{-3}$, where the nuclei
can be nonspherical and form clusters
\cite{pr95}.

Matter of the outer NS core consists
of neutrons, with admixture of protons, electrons
and possibly muons. This core may be about several km thick
and extends to $\rho \la 2 \rho_0$.
The inner core occupies deeper central layers
creating the main mystery of NSs since its composition
cannot be determined
uniquely by the present theories. 
There are many theoretical models predicting 
appearance of hyperons, pion or kaon condensates,
or quark matter.

\section{Inner crust matter}

Let us adopt the
ground-state 
model of crustal matter \cite{nv73,oyamatsu93} and
focus on the
temperature range from $\sim 10^8$ to $\sim 10^9$ K
of interest for NS cooling (Sect.\ 3).
The nuclear composition does not depend on $T$ for
$T \la 3 \times 10^9$ K. The nuclear charge number
is $Z \sim$40--50. 
The number of neutrons per a Wigner--Seitz cell
can reach $\sim 10^3$ at $\rho \sim 10^{13}$
g cm$^{-3}$; they mainly belong to
the neutron liquid outside the nuclei. At $\rho \sim 10^{14}$
g cm$^{-3}$ the neutron and proton
density distributions within the nucleus become smooth and
the nuclear radius becomes $\sim \, 0.5 a$, where $a$ is
the radius of the Wigner--Seitz cell.  
For $T=10^9$ K at this $\rho$
the Coulomb coupling parameter $\Gamma = (Ze)^2/(aT)$ reaches
$\sim 10^3$, and $T/T_p$ is $ \sim 0.3$,
where $T_p$ is the ion plasma temperature.
Accordingly the Coulomb crystal is nearly classical,
although it becomes quantum at lower $T$. 

The equation of state (EOS) in the inner crust 
is almost temperature independent.
The pressure is mainly determined by 
electrons at $\rho \sim \rho_{\rm d}$ and by neutrons 
at $\rho \sim \rho_{cc}$.
The neutron drip  
greatly softens the EOS at $\rho > \rho_{\rm d}$,
but the neutron liquid introduces considerable stiffness at $\rho \ga 10^{13}$
g cm$^{-3}$. 

If the neutrons were nonsuperfluid they would determine
the heat capacity in the inner crust.
The superfluidity
can greatly reduce the neutron contribution,
making the heat capacity of Coulomb crystal \cite{ph73}
dominant, for given temperatures.
 
The neutrino emissivity in the crust is produced by
several mechanisms. The most important are neutrino
pair bremsstrahlung due to scattering of electrons
off nuclei (e.g.\ ref.\ \cite{kppty99})
and plasmon decay into neutrino pair.

Finally, the thermal conductivity in the inner crust
is mainly provided by electrons which scatter off
atomic nuclei \cite{gyp00}. It 
depends weakly on $T$ for
$T=10^8$--$10^9$ K as shown in fig.\ 1a where
we use smooth--composition model \cite{kppty99}
of spherical nuclei \cite{nv73, oyamatsu93} 
at $\rho < 10^{14}$ g cm$^{-3}$ and its extrapolation
to higher $\rho$ in the crust.
It is important that 
at $\rho \ga \rho_{cc}/10$ the conductivity
is sensitive to the size of the proton charge distribution
within the nuclei. For instance, calculation
of the conductivity for pointlike nuclei 
at $\rho \sim 10^{14}$ g cm$^{-3}$ underestimates
the conductivity by a factor of 3--5. The conductivity
in the core \cite{gy95} is  $ \sim 10^2$ times higher
(fig.\ 1a) since there are no such efficient
electron scatterers as atomic nuclei there.

\section{Cooling of young neutron stars}

NSs are born very hot in supernova explosions, 
with the internal temperature $T \sim 10^{11}$, but gradually cool down.
We have calculated a number of cooling curves,
the surface temperatures $T_s^\infty$ as detected by
a distant observer versus stellar age $t$.
The detailed description
of the results is given elsewhere \cite{gyp00}.

We have used the code which calculates NS cooling by
solving the equations of heat conduction within the NS
taking into account
neutrino energy losses from the NS interior
and photon emission from the surface. The effects of General
Relativity are included explicitly.
We have adopted the same
EOS in the NS core (composed of neutrons, protons and
electrons) as in ref.\ \cite{pa92}. The code includes all
relevant sources of neutrino energy losses,
heat capacity and thermal conductivity in the core and crust.
The effects of neutron superfluidity in the core and
crust and proton superfluidity in the core have been
incorporated to test various theoretical models of superfluidity.

Although the inner crust is hidden deeply within the NS,
the cooling is sensitive to its properties
at the early cooling stage ($t \la 10$--300 yr) as long as
the internal thermal relaxation is not
established.
The NS energy losses at this stage are mainly provided 
by neutrino emission.
The lower
thermal conductivity in the crust
delays the crustal thermal relaxation making
it very pronounced in the cooling curves.

\begin{figure}[t!]
\begin{center}
\leavevmode
\epsfysize=8cm 
\epsfbox[20 20 540 340]{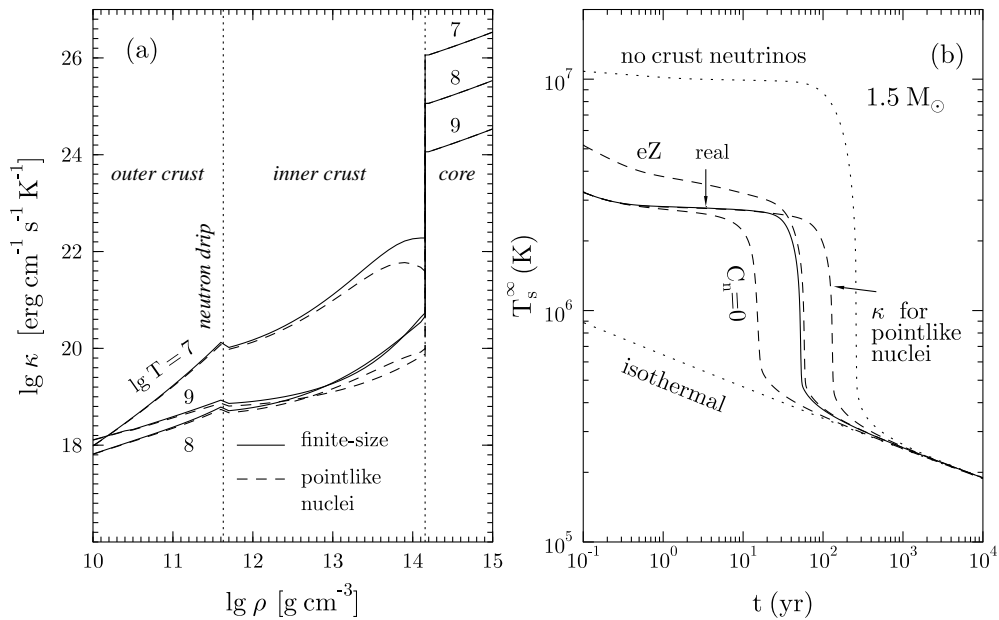}
\end{center}
\caption[ ]{ 
  (a) Electron thermal conductivity in NS crust and core
  at three values of $T$ for finite-size and pointlike
  nuclei in the crust. (b) Decrease of surface
  temperature $T_s^\infty$ of a young NS for 
  several NS crust models.
}
\label{fig1}
\end{figure}

To test this statement we have run several 
models switching on and off various ingredients. 
Fig.\ 1b shows some cooling curves for a 
nonsuperfluid NS model with $M=1.5 M_\odot$ and $R=11.38$ km.
The central density $1.42 \times 10^{15}$ g cm$^{-3}$
is high enough to allow direct Urca process,
the most powerful neutrino emission mechanism,
to operate in a central NS kernel. This leads to fast neutrino
cooling of the NS core (e.g,
refs.\ \cite{pa92,yls99}). However the surface temperature
remains independent of the core temperature at the initial
nonrelaxed stage.
The solid line is the real cooling curve.
The end of the relaxation stage manifests itself in
a spectacular drop of $T_s^\infty$ by about
one order of magnitude. 
The relaxation time, 
defined  \cite{lvpp94} as the moment of the steepest fall
of $T_s^\infty(t)$, is $t_r \approx 52$ yr, for the real cooling.
Switching off all neutrino emission
in the crust (the upper dotted line) would
delay the relaxation to 258 yr keeping extremely
high surface temperature, $T_s^\infty \sim 10^7$ K, at the nonrelaxed stage. 
Turning on the neutrino
bremsstrahlung ($eZ$) alone would lead to $t_r$=54 yr,
close to the real value. Restoring full neutrino emission
but switching off the heat capacity of neutrons ($C_n=0$) in the
crust (imitating thus the effect of strong neutron superfluidity)
would speed up the relaxation
to 15 yr. If we additionally turned off
the heat capacity of nuclei the relaxation would
speed up further to about 11 yr. On the other hand, if we 
restored the full heat capacity but ignored
quantum suppression of the heat capacity of nuclei
at $T \ll T_p$, the latter heat capacity would become important
in older NSs, $t \ga 10^4$ yr, strongly delaying the cooling. 
Finally, switching on the
heat capacity and neutrino emission but assuming infinite
thermal conductivity
(lower dotted line) would remove the
relaxation stage at all and lead to the fast drop of $T_s^\infty$
in a young NS.
The relaxation time is very sensitive to the values of
the thermal conductivity $\kappa$ in the crust at $\rho \sim 10^{14}$
g cm$^{-3}$. For instance, using the lower conductivity
for pointlike nuclei (fig.\ 1a) would delay
the relaxation to 130 yr.

Our calculations confirm the scaling relation \cite{lvpp94},
$t_r = \alpha \tau$, where $\alpha = (\Delta R /1~{\rm km})^2 
(1-r_g/R)^{-3/2}$ is the factor which depends on
the crust thickness $\Delta R=R-R_{cc}$, NS mass
and radius [$r_g=2GM/(c^2R)$ being the gravitational
radius], while $\tau$ depends solely on physical
properties of crustal matter. For the NS model in fig.\ 1b
we have $\Delta R=0.93$ km and $\alpha=1.81$. 
The scaling enables one to calculate
$t_r$ for other NS models.

It is well known that the NS cooling theory
can be used for interpretation of observations 
of middle-aged NSs ($t \sim 10^4$--$10^6$ yrs)
providing viable information on physical
properties of matter in the NS cores (e.g., ref.\ \cite{yls99}).
Now we see that cooling of young NSs, $t \la 100$ yr, 
depends strongly on
the thermal conductivity, heat capacity, and neutrino emissivity
of the Coulomb--nuclear plasma in the inner NS crusts,
at $\rho \sim 10^{14}$ g cm$^{-3}$. This gives potentially
powerful method to test theoretical predictions of 
the properties of such plasma,
particularly, sizes of highly unusual atomic nuclei
and critical temperatures of the neutron superfluidity.
Unfortunately,
such young NSs have not been detected so far.
Hopefully they will be observed in the near future
in not too distant supernova explosions.
This will enable one to realize the above method
in practice.

\begin{acknowledgements}
 The work was
supported in part by RFBR (grant No.\ 99-02-18099) and INTAS (grant 
No.\ 96-0542). DGY and AYP are grateful to DFG for support provided to
attend PNP10.
\end{acknowledgements}

\vspace*{-1cm}

\begin{received}
Received 1 October 2000
\end{received}

\end{document}